# Unified model for breathing solitons in fibre lasers: Mechanisms across below- and above-threshold regimes

Ying Zhang, [1,†] Bo Yuan, [1,†] Junsong Peng, [1,2,3, *] Xiuqi Wu,[1] Yulin Sheng, [1] Yuxuan Ren, [1] Christophe Finot, [4] Sonia Boscolo, [5,6,7] Heping Zeng [1,3, **]

[1]State Key Laboratory of Precision Spectroscopy, and Hainan Institute, East China Normal University, Shanghai 200062, China
[2]Collaborative Innovation Center of Extreme Optics, Shanxi University, Taiyuan, Shanxi 030006, China
[3]Chongqing Key Laboratory of Precision Optics, Chongqing Institute of East China Normal University, Chongqing 401120, China
[4]Université Bourgogne Europe, CNRS, Laboratoire Interdisciplinaire Carnot de Bourgogne (ICB), UMR 6303, F-21000 Dijon, France
[5]VPIphotonics GmbH, 10587 Berlin, Germany
[6]Aston Institute of Photonic Technologies, Aston University, Birmingham B4 7ET, United Kingdom
[7]Istituto Nazionale di Fisica Nucleare, Laboratori Nazionali del Sud (INFN-LNS), 95125 Catania, Italy
† These authors contributed equally to this work
*jspeng@lps.ecnu.edu.cn ** hpzeng@phy.ecnu.edu.cn

**Abstract:** The emergence of breathing solitons in mode-locked lasers presents a fundamental challenge for the theoretical modelling of mode locking, with the mechanisms underlying below- and above-threshold breathing solitons, and the origins of their distinct nonlinear dynamics, remaining poorly understood. Here, we develop a model that incorporates both spatial and temporal gain dynamics, enabling us to elucidate the origins of these two classes of pulsating states. We show that below-threshold breathing solitons arise from the interplay between Q-switching and soliton shaping, whereas Kerr nonlinearity and dispersion dominate the formation of above-threshold breathers. The model further captures the markedly different dynamical properties of these regimes. Experimental observations corroborate the simulations, validating the predictive power of the framework. Beyond providing a refined theoretical basis for ultrafast laser design, this work advances the broader understanding of non-equilibrium dynamics in mode-locked lasers and offers new perspectives on breathing soliton phenomena across diverse physical systems.

Mode-locked fibre lasers, renowned for their ability to generate ultrashort pulses, are indispensable tools in physics, chemistry, and materials science [1, 2]. Beyond stationary soliton emission, these systems can also sustain breathing, or pulsating, solitons [3-12], which undergo periodic oscillations in energy over successive cavity roundtrips. Recent studies have established that such dissipative structures are intimately connected to synchronisation phenomena [13], positioning fibre lasers as a powerful experimental platform for probing dynamical behaviors typically associated with nonlinearly coupled oscillators [14]. In breathing-soliton lasers, two intrinsic frequencies naturally arise: the pulse repetition rate ($f_r$), determined by the cavity length, and the breathing frequency ($f_b$) of the pulse train. Their ratio ($f_b/f_r$), referred to as the winding number, can exhibit a universal fractal organisation akin to the "devil's staircase" widely recognised in synchronisation theory [8]. Under certain conditions, a third frequency may emerge, signaling the onset of modulated subharmonics [10], which can ultimately drive the system into chaos via a distinct pathway—the modulated-subharmonic route [15]. Furthermore, in breathing-soliton lasers the synchronisation regions deviate from the classical Arnold tongues due to the pronounced disparity in strength between the internal frequencies, and can also display "holes" within them [16, 17].

The frequency locking and fractal dynamics described above are observed exclusively in lasers operating near zero net dispersion—whether slightly anomalous or normal—where breathing solitons emerge at pump powers exceeding the stability limit of stationary solitons. These states are therefore referred to as *above-threshold* breathing solitons. In contrast, in cavities with relatively strong net normal dispersion, breathing solitons appear below the pump threshold required for conventional soliton mode locking [5]. Beyond their distinct excitation conditions, these two types of breathing solitons exhibit markedly different experimental behaviours: i) Oscillation period — Above-threshold breathers oscillate with short periods spanning only a few cavity roundtrips, whereas below-threshold breathers display much longer periods, extending over hundreds to thousands of roundtrips. ii) Synchronisation capability — The high breathing frequency in the above-threshold regime promotes frequency locking, yielding comb-like radiofrequency (RF) spectra and enabling the emergence of high-order frequency-locked states [8, 18, 19]. By contrast, in the below-threshold case, the much slower breathing frequency produces RF spectra densely clustered around $f_r$, lacking strict commensurability with the cavity roundtrip time. iii) Optical spectra — Sidebands are a characteristic feature of the above-threshold regime but are absent for below-threshold breathing solitons.

Understanding the origins of these differences is of fundamental importance to ultrafast laser dynamics, soliton physics, and nonlinear science. Nevertheless, progress has been hindered by the absence of a universal theoretical framework capable of capturing both types of breathing solitons. The time-varying dynamics of breathing solitons has long posed a challenge for modelling mode locking, and two distinct approaches are commonly employed to obtain quantitative or qualitative insights into the two regimes. The lumped-cavity approach, based on the generalised nonlinear Schrödinger equation (GNLSE) with a simplified gain term and explicit treatment of each cavity component, has proven effective in quantitatively predicting the behaviour of above-threshold breathing solitons [6, 8, 10, 16]. However, it cannot describe the generation of below-threshold breathers. For the latter, an averaged model in the form of a master equation, namely the complex Ginzburg–Landau equation (CGLE), has offered valuable qualitative understanding [4, 5, 20]. Yet, because the CGLE averages over the discrete action of the individual cavity components, it cannot fully capture the underlying generation mechanisms, and the parameters of the equation are not straightforwardly linked to the physical properties of specific cavity segments.

A universal theoretical framework capable of simultaneously describing both below- and above-threshold breathing solitons is therefore highly desirable, as it would allow the distinct generation mechanisms and the origins of their markedly different dynamics to be unambiguously revealed. Understanding the physical origins of these dynamics can guide the mitigation of these pulsating instabilities (as discussed later), thereby expanding the range of stable mode-locking regimes—a key consideration in laser design. Furthermore, the unified model can be applied to explore emerging gain-related non-equilibrium dynamics in mode-locked lasers, including turn-on dynamics [21, 22], replica symmetry breaking [23, 24], resonant excitation of soliton molecules [25], shock waves[5], and many other phenomena [26-31].

Here, we introduce a modified discrete mode-locked laser model that explicitly accounts for the spatiotemporal dynamics of the gain medium. Although recent studies have introduced a spatially varying population inversion, they typically assume time-independent inversion [32-35]. Here, we revise the standard discrete model by incorporating the slow gain dynamics that the CGLE captures through its coefficients, while preserving the fast intracavity mapping characteristic of lumped cavity models. With this model, we are able to reproduce the generation of both below- and above-threshold breathing solitons along with their characteristic signatures, providing insight into the mechanisms responsible for the pronounced differences between the two regimes. Similar approaches have been recently employed to simulate breather molecules in synchronised mode-locked lasers [9] and describe bifurcations in nonlinear-loop-mirror lasers [36]. In parallel, spatiotemporal gain dynamics have also been incorporated into models of multimode fibre amplifiers, enabling rigorous treatment of key physical processes such as differential modal gain, gain-managed nonlinear evolution, and Kerr beam self-cleaning [37]. These advances provide essential physical insight and constitute a necessary foundation for the future development of quantitatively reliable models of the complex spatiotemporal nonlinear dynamics governing mode-locked multimode fibre lasers [38].

Two fibre lasers with distinct dispersion values are employed. The first operates in a near-zero dispersion regime (~0.007 ps²) to support above-threshold breathing solitons, while the second incorporates strong net normal dispersion (~0.026 ps²) to generate below-threshold breathers. Figure 1 illustrates the near-zero dispersion laser. The gain medium (erbium-doped fibre, EDF) is pumped by a laser diode via a wavelength-division multiplexer, and an isolator ensures unidirectional operation. Nonlinear polarisation rotation (NPR) serves as the mode-locking mechanism, implemented using a quarter-wave plate, a half-wave plate, and a polarisation beam splitter, which simultaneously functions as the output coupler. The high net normal dispersion cavity follows a similar architecture but incorporates an additional normal-dispersion fibre (NDF) to achieve the target dispersion. The detailed properties of all components, used as input parameters for the simulations, are provided in Tables 1 and 2 of the Supplemental Materials (SMs) [39]. For experimental characterisation, real-time optical spectra are measured using the time-stretch technique [40-42], in which a long NDF temporally separates different frequency components, allowing spectral reconstruction on the oscilloscope. RF characteristics are evaluated with an electrical spectrum analyser.

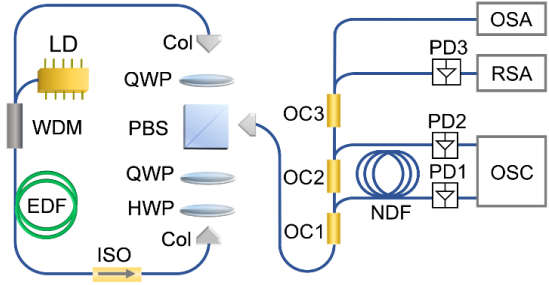

Fig. 1 The laser setup. EDF: erbium-doped fibre, WDM: wavelength division multiplexer, Col: collimator, QWP: quarter-wave plate, PBS: polarisation beam splitter, HWP: half-wave plate, ISO: isolator, OC: optical coupler, NDF: normal-dispersion fibre, PD: photodetector, OSC: oscilloscope, ESA: electrical spectrum analyser, OSA: optical spectrum analyser.

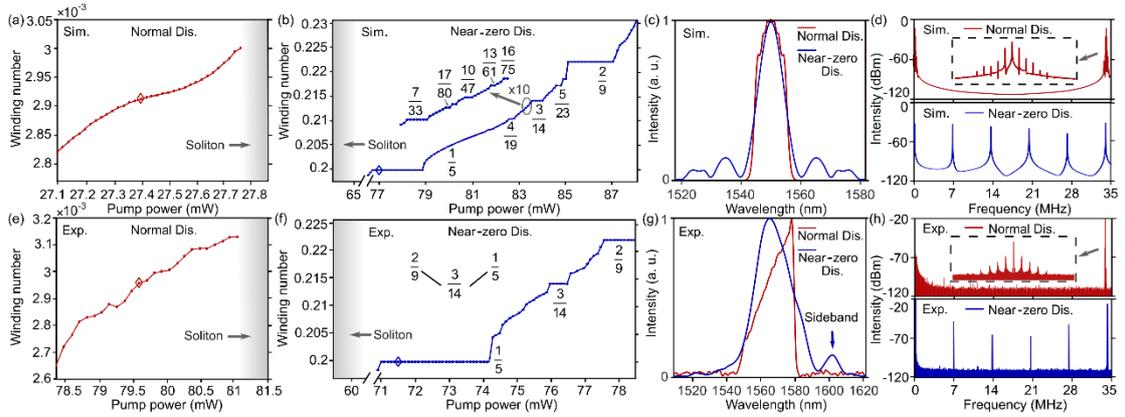

Fig. 2 Distinct features of breathing solitons in normal- and near-zero-dispersion cavities. (a, b) Simulated winding numbers as a function of pump power. Representative (c) optical spectra and corresponding (d) RF signals for the two classes of breathing solitons recorded at pump powers indicated by the diamonds in (a, b). (e-h) Corresponding experimental results. The normal- and near-zero-dispersion cavities have similar repetition rates ($f_r$) of 34.1 and 34.2 MHz, respectively. The shaded areas in (a, b) and (e, f) indicate the soliton regimes. The inset in (b) shows the emergence of additional plateaus when the pump-power resolution is increased by an order of magnitude, highlighting the fractal nature of the winding-number distribution. Insets in (d) and (h) display magnified regions around $f_r$.

The pulse evolution within each fibre segment is described by the GNLSE, which in the scalar approximation takes the form [43]:

$$\frac{\partial \psi}{\partial z} = -\frac{i\beta_2}{2}\frac{\partial^2 \psi}{\partial \tau^2} + i\gamma |\psi|^2 \psi + \frac{g}{2}\left(\psi + \frac{1}{\Omega^2}\frac{\partial^2 \psi}{\partial \tau^2}\right), \qquad (1)$$

where $\psi(z,\tau)$ denotes the slowly varying electric field envelope, $\tau$ is the retarded time, and $z$ is the longitudinal propagation coordinate. The parameters $\beta_2$ and $\gamma$ represent the second-order group-velocity dispersion and Kerr nonlinearity, respectively, while $g$ and $\Omega$ correspond to the gain amplitude and bandwidth in the active fibre. The NPR-based mode-locking mechanism is modelled via an instantaneous, monotonically increasing nonlinear transfer function applied to the field amplitude: $T = \sqrt{R_0 + \Delta R \cdot |\psi|^2/(P_{\text{sat}} + |\psi|^2)}$, where $R_0$ is the unsaturated transmittance, $\Delta R$ is the modulation depth, and $P_{\text{sat}}$ is the saturation power. In standard mode-locked laser models, the gain amplitude is often approximated by a simplified saturable function [44]. While this approach is sufficient to reproduce above-threshold breather excitation, it fails to capture the generation of below-threshold breathers (see Fig. S1 in SMs). This limitation highlights the need for an improved description of the gain dynamics, since the conventional model neglects the realistic spatiotemporal evolution of the population inversion in the gain fibre. Accordingly, the gain dynamics is determined by the two-level rate equations governing the distribution of the signal and pump powers, $P_s(z)$ and $P_p(z)$, along the gain fibre [45]:

$$\frac{dP_s}{dz} = \Gamma_s(\sigma_s^e N_2 - \sigma_s^a N_1)P_s, \tag{2}$$

$$\frac{dP_p}{dz} = \Gamma_p(\sigma_p^e N_2 - \sigma_p^a N_1)P_p, \tag{3}$$

$$\frac{\partial N_2}{\partial t} = -\frac{N_2}{\tau_2} - \frac{P_p \lambda_p}{A_{co} hc}(\sigma_p^e N_2 - \sigma_p^a N_1) - \frac{P_s \lambda_s}{A_{co} hc}(\sigma_s^e N_2 - \sigma_s^a N_1), \tag{4}$$

where $\sigma_{s,p}^{e/a}$ are the emission and absorption cross sections at the signal ($\lambda_s$=1550 nm) and pump ($\lambda_p$=980 nm) wavelengths, respectively. $N_1$ and $N_2$ denote the ground- and excited-state population densities, with $N = N_1 + N_2$ constant, $\tau_2$ is the excited-state lifetime, $A_{co}$ is the core area of the fibre, and $h$ and $c$ are Planck's constant and the speed of light in vacuum, respectively. The solution of Eqs. (2)-(4) yields the gain coefficient along the active fibre during the $m$-th cavity roundtrip: $g(z;m) = d(\ln P_s)/dz$, which is then used as the initial condition for computing the gain coefficient in the subsequent roundtrip, $m$+1. Thus, only slow gain dynamics is considered, while the fast gain variation across a pulse is neglected. Eq. (1) is integrated using a symmetric split-step Fourier method, and Eqs. (2)–(4) are solved with a fourth-order Runge–Kutta scheme.

Remarkably, the lumped model incorporating the spatiotemporal dependence of the gain successfully simulates breathing-soliton generation in both net-normal and near-zero dispersion regimes. The main simulation results are summarised in panels (a–d) of Fig. 2. Figures 2(a) and 2(b) show the scaling of the winding number ($f_b/f_r$) with pump power in the normal- and near-zero dispersion regimes, respectively, with the shaded regions indicating stationary soliton regimes. Figures 2(c) and 2(d) present representative optical spectra and RF signals in the two regimes. The results reveal markedly distinct behaviours between the two regimes. In the net-normal dispersion regime, breathing solitons emerge *below* the pump threshold for stable solitons, whereas in the near-zero dispersion regime, they appear *above* threshold. The winding number differs by nearly two orders of magnitude, corresponding to breathing periods of hundreds of cavity roundtrips for the normal-dispersion regime versus only a few roundtrips in the near-zero dispersion case. This difference in winding number directly influences synchronisation, as the locking width scales with the winding number [14, 46]. Notably, the winding number curve in Fig. 2(b) exhibits multiple plateaus following the Farey sum ($\frac{1}{5} \oplus \frac{2}{9} = \frac{3}{14}$, highlighted as one example in the inset), reflecting the universal fractal behaviour of frequency locking, known as the devil's staircase [8, 47-50]. In contrast, no frequency locking is observed in Fig. 2(a). These differences are also evident in the RF spectra [Fig. 2(d)]: the above-threshold frequency-locked breathers (blue) display a comb structure, with the dominant line corresponding to the breathing frequency, whereas the RF spectrum of below-threshold breathers (red) is densely populated near the cavity repetition frequency, and the breathing sidebands do not appear at exact subharmonics of the repetition rate. Finally, the optical spectra [Fig. 2(c)] reveal that above-threshold breathing solitons exhibit characteristic sidebands, whereas below-threshold breathers do not.

Importantly, the model produces all key features of the two experimentally observed breather types [Fig. 2(e–h)]. Discrepancies in the extent and symmetry of the optical spectra are attributed to residual higher-order dispersion and polarisation effects not fully captured in the scalar, lumped model.

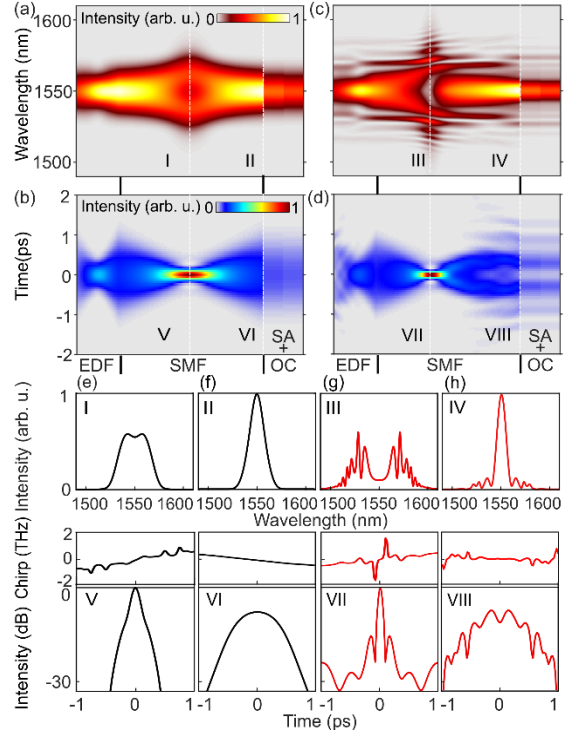

Fig. 3 Simulated intracavity dynamics of solitons and above-threshold breathing solitons. (a, c) Spectral and (b, d) temporal evolutions of solitons [panels (a, b)] and breathing solitons [panels (c, d)] at pump power of 30 and 77 mW, respectively. (e, f) Spectrum (top), chirp (middle), and temporal intensity (bottom) profiles of solitons at the midpoint and end of the SMF, as indicated by the dashed lines in (a, b). (g, h) Corresponding spectral and temporal characteristics of breathing solitons at the positions marked by dashed lines in (c, d).

In the following, we analyse the distinct generation mechanisms underlying the two types of breathing solitons. Figures 3(a–d) compare representative intracavity temporal and spectral evolutions of solitons and above-threshold breathing solitons (winding number 1/5) in the near-zero-dispersion regime. Both solitons [Figs. 3(a, b)] and breathing solitons [Figs. 3(c, d)] exhibit pronounced intracavity variations across the two fibre segments with opposite dispersion. The spectrum broadens to its maximum and develops sidebands when the pulse reaches its minimum temporal duration near the midpoint of the standard single-mode fibre (SMF). Cross-sections of the broadest spectrum and its corresponding temporal intensity (with chirp profiles), marked by dashed lines in Figs. 3(a, b), are shown in Fig. 3(e). Notably, the broadened spectrum narrows rapidly in the remaining SMF [Fig. 3(a)], a phenomenon attributed to chirp compensation in passive fibres [51]. Here, the positive chirp induced by the normal-dispersion EDF and Kerr nonlinearity [Fig. 3(e), middle] is counteracted by the anomalous-dispersion SMF's negative chirp [Fig. 3(f), middle], redistributing energy from the pulse wings to the center and narrowing the spectrum [Fig. 3(f), top]. This spectral narrowing is essential for solitons to satisfy the cavity boundary conditions and remain self-consistent over a single round trip.

At higher pump powers, increased pulse energy amplifies the Kerr effect, producing significantly more complex temporal and spectral profiles for breathing solitons [Fig. 3(g)] through enhanced Kerr–dispersion interplay. The resulting chirp profile also exhibits intricate structure [Fig. 3(g), middle]. Unlike solitons, whose weaker nonlinear chirp can be partially compensated by the SMF's linear chirp from second-order dispersion [52], breathing solitons maintain a strong nonlinear chirp [Fig. 3(h), middle] that resists full compensation, leaving pronounced spectral and temporal wings [Fig. 3(h), top and bottom, respectively]. As a result, the pulse cannot stabilise within a single round trip but instead repeats periodically over multiple round trips. Meanwhile, the excited-state population density ($N_2$) also varies periodicity, which cannot be revealed by the standard discrete model (see Fig. S2 of SMs).

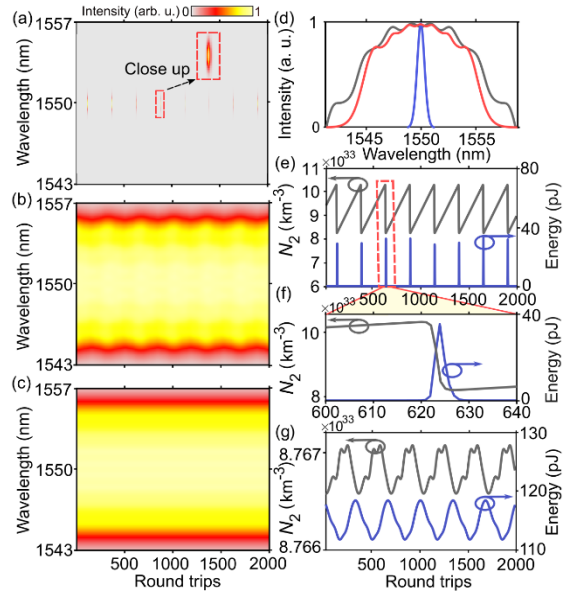

Fig. 4 Simulated Q-switched pulses and below-threshold breathing solitons. (a–c) Roundtrip-resolved spectral evolution of Q-switched pulses, below-threshold breathing solitons, and stable solitons, respectively, at pump powers of 1, 27.4, and 28.5 mW. (d) Representative optical spectra of Q-switched pulses (blue), breathing solitons (red), and stable solitons (black). (e, g) Roundtrip-resolved evolution of pulse energy (blue) and excited-state population $N_2$ (black) for Q-switched pulses and breathing solitons, respectively. (f) Magnified view of the boxed region in (e), highlighting the clear anti-phase correlation between energy and $N_2$.

In contrast, the above pulse-shaping dynamics requiring anomalous-dispersion SMF do not contribute to the generation of below-threshold breathing solitons, as strong normal dispersion inherently suppresses this effect. Q-switching likely plays a role, consistent with their appearance below the mode-locking threshold, as confirmed in Fig. 4. By incorporating the rate equations, Q-switching is first simulated within the discrete mode-locked laser model. Figure 4(a) shows the periodic evolution of the optical spectrum of Q-switched pulses over consecutive round trips (experimental Q-switching results are provided in Fig. S3 of SMs). As the pump power increases, the pulse spectrum broadens due to the Kerr effect while retaining its oscillatory character [Fig. 4(b)]. At higher pump powers, stable solitons emerge, exhibiting an invariant pulse spectrum across round trips [Fig. 4(c)]. Figure 4(d) compares the representative optical spectra of all three states, showing that the Q-switched state has a significantly narrower spectrum, as the Kerr effect is negligible. Figure 4(e) presents Q-switching-generated energy bursts (blue) and the corresponding $N_2$ evolution (black), revealing anti-phase dynamics further resolved in Fig. 4(f). Notably, this energy–$N_2$ phase relationship persists in below-threshold breathing solitons [Fig. 4(g)], providing direct evidence that these pulses originate from Q-switching modulation. According to laser physics, the onset of Q-switching is determined by the pulse energy and the parameters of the saturable absorber [53]. Indeed, Fig. 4(a–c) illustrates that the appearance of Q-switching depends on the pump power. Further simulations confirm that the parameters of the saturable absorber can also control the onset of Q-switching (see Fig. S4 in SMs).

An important question is whether the transition between below- and above-threshold breathers is continuous when tuning the dispersion. Further simulations show a sharp transition (see Fig. S5 in SMs).

The standard discrete model fails to capture many essential laser dynamics that emerge when pulse propagation is coupled with realistic gain effects. In contrast, our modified model provides a more comprehensive framework for investigating these complex nonlinear phenomena. For example, although Q-switching plays a critical role in the buildup of stationary solitons—as confirmed experimentally [22, 54]—the standard model only partially reproduces this process [55]. Our model successfully captures these dynamics, as illustrated in Fig. S6 of SMs.

In conclusion, our discrete mode-locked laser model reveals fundamentally distinct features and generation mechanisms for below- and above-threshold breathing solitons. These insights have significant implications for ultrafast laser design. Since stationary solitons are generally preferred for practical applications, our results suggest strategies to suppress breathing solitons and thereby expand the parameter space for stable operation. Specifically, below-threshold breathers can be suppressed by reducing Q-switching—either through the use of a saturable absorber with a small modulation depth [53] or by increasing cavity dispersion to increase pulse energy. Above-threshold breathers can be mitigated by minimising the length of anomalous-dispersion fibre.

Beyond these immediate design implications, our results also bear on the broader problem of nonlinear dynamical complexity in mode-locked lasers. Recent research has increasingly focused on low-dimensional soliton chaos [15, 56-59]. Owing to the coupling of four governing equations, our model provides a natural framework for exploring higher-dimensional dynamical regimes, including hyperchaos [60-62].

**Acknowledgement**

We acknowledge support from the Innovation Program for Quantum Science and Technology (2023ZD0301000), National Natural Science Fund of China (12434018, 62475073, 1243000542, 11621404, 11561121003, 11727812, 61775059, 12074122,

**References**
[1]. U. Keller, "Recent developments in compact ultrafast lasers," Nature **2003**, 424, 831-838.
[2]. M. E. Fermann, and I. Hartl, "Ultrafast fibre lasers," Nature Photonics **2013**, 7, 868-874.
[3]. J. M. Soto-Crespo, N. Akhmediev, and A. Ankiewicz, "Pulsating, creeping, and erupting solitons in dissipative systems," Phys. Rev. Lett. **2000**, 85, 2937-2940.
[4]. J. M. Soto-Crespo, M. Grapinet, P. Grelu, and N. Akhmediev, "Bifurcations and multiple-period soliton pulsations in a passively mode-locked fiber laser," Physical Review E **2004**, 70, 066612.
[5]. J. Peng, S. Boscolo, Z. Zhao, and H. Zeng, "Breathing dissipative solitons in mode-locked fiber lasers," Science Advances **2019**, 5, eaax1110.
[6]. T. Xian, L. Zhan, W. Wang, and W. Zhang, "Subharmonic Entrainment Breather Solitons in Ultrafast Lasers," Physical Review Letters **2020**, 125, 163901.
[7]. J. Guo, S. Cundiff, J. Soto-Crespo, and N. Akhmediev, "Concurrent Passive Mode-Locked and Self-Q-Switched Operation in Laser Systems," Physical Review Letters **2021**, 126, 224101.
[8]. X. Wu, Y. Zhang, J. Peng, S. Boscolo, C. Finot, and H. Zeng, "Farey tree and devil's staircase of frequency-locked breathers in ultrafast lasers," Nature Communications **2022**, 13, 5784.
[9]. Y. Cui, Y. Zhang, L. Huang, A. Zhang, Z. Liu, C. Kuang, C. Tao, D. Chen, X. Liu, and B. A. Malomed, "Dichromatic "breather molecules" in a mode-locked fiber laser," Physical Review Letters **2023**, 130, 153801.
[10]. X. Wu, J. Peng, S. Boscolo, C. Finot, and H. Zeng, "Synchronization, Desynchronization, and Intermediate Regime of Breathing Solitons and Soliton Molecules in a Laser Cavity," Physical Review Letters **2023**, 131, 263802.
[11]. X. Wu, J. Peng, S. Boscolo, Y. Zhang, C. Finot, and H. Zeng, "Intelligent Breathing Soliton Generation in Ultrafast Fiber Lasers," Laser & Photonics Reviews **2022**, 16, 2100191.
[12]. Z. Wang, A. Coillet, S. Hamdi, Z. Zhang, and P. Grelu, "Spectral Pulsations of Dissipative Solitons in Ultrafast Fiber Lasers: Period Doubling and Beyond," Laser & Photonics Reviews, n/a, 2200298.
[13]. Junsong Peng, Xiuqi Wu, Huiyu Kang, Anran Zhou, Ying Zhang, Heping Zeng,Christophe Finot & Sonia Boscolo, "Nonlinear dynamics in breathing-soliton lasers,"Advances in Physics: X **2025**, 10, 2580628.
[14]. A. Pikovsky, M. Rosenblum, and J. Kurths, "Synchronization: a universal concept in nonlinear science," Cambridge University Press, 2001.
[15]. H. Kang, A. Zhou, Y. Zhang, X. Wu, B. Yuan, J. Peng, C. Finot, S. Boscolo, and H. Zeng, "Observation of optical chaotic solitons and modulated subharmonic route to chaos in mode-locked laser," Physical Review Letters **2024**, 133, 263801.
[16]. X. Wu, J. Peng, B. Yuan, S. Boscolo, C. Finot, and H. Zeng, "Unveiling the complexity of Arnold's tongues in a breathing-soliton laser," Science Advances **2025**, 11, eads3660.
[17]. G. V. Papaioannou, D. K. Yue, M. S. Triantafyllou, and G. E. Karniadakis, "Evidence of holes in the Arnold tongues of flow past two oscillating cylinders," Physical review letters **2006**, 96, 014501.

[18]. Y. Ni, Z. Xie, Y. Sheng, J. Ge, J. Peng, and H. Zeng, "Revealing fractal dynamics by dispersion tuning in a mode-locked fiber laser," J. Opt. Soc. Am. B **2024**, 41, 2285-2289.
[19]. Y. Zhang, X. Wu, J. Peng, and H. Zeng, "On the universality of fractal breathers in mode-locked fibre lasers," Optics Communications **2023**, 547, 129845.
[20]. W. Chang, J. M. Soto-Crespo, P. Vouzas, and N. Akhmediev, "Extreme soliton pulsations in dissipative systems," Physical Review E **2015**, 92, 022926.
[21]. G. Herink, B. Jalali, C. Ropers, and D. Solli, "Resolving the build-up of femtosecond mode-locking with single-shot spectroscopy at 90 MHz frame rate," Nat. Photon. **2016**, 10, 321-326.
[22]. P. Ryczkowski, M. Närhi, C. Billet, J. M. Merolla, G. Genty, and J. M. Dudley, "Real-time full-field characterization of transient dissipative soliton dynamics in a mode-locked laser," Nature Photonics **2018**.
[23]. N. P. Alves, W. F. Alves, A. C. Siqueira, N. L. Matias, A. S. Gomes, E. P. Raposo, and M. H. de Miranda, "Observation of Replica Symmetry Breaking in Standard Mode-Locked Fiber Laser," Physical Review Letters **2024**, 132, 093801.
[24]. A. L. Moura, P. I. R. Pincheira, A. S. Reyna, E. P. Raposo, A. S. L. Gomes, and C. B. de Araújo, "Replica Symmetry Breaking in the Photonic Ferromagneticlike Spontaneous Mode-Locking Phase of a Multimode Nd:YAG Laser," Physical Review Letters **2017**, 119, 163902.
[25]. F. Kurtz, C. Ropers, and G. Herink, "Resonant excitation and all-optical switching of femtosecond soliton molecules," Nature Photonics **2020**, 14, 9-13.
[26]. J. M. Dudley, G. Genty, A. Mussot, A. Chabchoub, and F. Dias, "Rogue waves and analogies in optics and oceanography," Nature Reviews Physics **2019**.
[27]. C. Lecaplain, P. Grelu, J. Soto-Crespo, and N. Akhmediev, "Dissipative rogue waves generated by chaotic pulse bunching in a mode-locked laser," Phys. Rev. Lett. **2012**, 108, 233901.
[28]. X. Wu, Y. Zhang, J. Peng, S. Boscolo, C. Finot, and H. Zeng, "Control of spectral extreme events in ultrafast fiber lasers by a genetic algorithm," Laser & Photonics Reviews **2024**, 18, 2200470.
[29]. S. Hamdi, A. Coillet, B. Cluzel, P. Grelu, and P. Colman, "Superlocalization reveals long-range synchronization of vibrating soliton molecules," Physical Review Letters **2022**, 128, 213902.
[30]. S. T. Cundiff, J. M. Soto-Crespo, and N. Akhmediev, "Experimental evidence for soliton explosions," Phys. Rev. Lett. **2002**, 88, 073903.
[31]. J. Peng, and H. Zeng, "Experimental Observations of Breathing Dissipative Soliton Explosions," Physical Review Applied **2019**, 12, 034052.
[32]. A. F. Runge, C. Aguergaray, R. Provo, M. Erkintalo, and N. G. Broderick, "All-normal dispersion fiber lasers mode-locked with a nonlinear amplifying loop mirror," Optical Fiber Technology **2014**, 20, 657-665.
[33]. A. F. Runge, N. G. Broderick, and M. Erkintalo, "Observation of soliton explosions in a passively mode-locked fiber laser," Optica **2015**, 2, 36-39.
[34]. Z. Wang, K. Nithyanandan, A. Coillet, P. Tchofo-Dinda, and P. Grelu, "Optical soliton molecular complexes in a passively mode-locked fibre laser," Nature communications **2019**, 10, 830.
[35]. S. Boscolo, C. Finot, I. Gukov, and S. K. Turitsyn, "Performance analysis of dual-pump nonlinear amplifying loop mirror mode-locked all-fibre laser," Laser Physics Letters **2019**, 16, 065105.

[36]. K. Krupa, T. M. Kardaś, and Y. Stepanenko, "Real-Time Observation of Double-Hopf Bifurcation in an Ultrafast All-PM Fiber Laser," Laser & Photonics Reviews **2022**, 16, 2100646.
[37]. Y.-H. Chen, H. Haig, Y. Wu, Z. Ziegler, and F. Wise, "Accurate modeling of ultrafast nonlinear pulse propagation in multimode gain fiber," J. Opt. Soc. Am. B **2023**, 40, 2633-2642.
[38]. L. G. Wright, D. N. Christodoulides, and F. W. Wise, "Spatiotemporal mode-locking in multimode fiber lasers," Science **2017**, 358, 94-97.
[39]. See Supplemental Material [url] for more information.
[40]. A. F. Runge, C. Aguergaray, N. G. Broderick, and M. Erkintalo, "Coherence and shot-to-shot spectral fluctuations in noise-like ultrafast fiber lasers," Opt. Lett. **2013**, 38, 4327-4330.
[41]. A. Mahjoubfar, D. V. Churkin, S. Barland, N. Broderick, S. K. Turitsyn, and B. Jalali, "Time stretch and its applications," Nature Photonics **2017**, 11, 341-351.
[42]. T. Godin, L. Sader, A. Khodadad Kashi, P.-H. Hanzard, A. Hideur, D. J. Moss, R. Morandotti, G. Genty, J. M. Dudley, and A. Pasquazi, "Recent advances on time-stretch dispersive Fourier transform and its applications," Advances in Physics: X **2022**, 7, 2067487.
[43]. H. A. Haus, "Mode-locking of lasers," IEEE J. Sel. Top. Quantum Electron. **2000**, 6, 1173-1185.
[44]. B. Oktem, C. Ulgudur, and F. O. Ilday, "Soliton-similariton fibre laser," Nature Photonics **2010**, 4, 307-311.
[45]. E. Desurvire, and M. N. Zervas, "Erbium-doped fiber amplifiers: principles and applications," (American Institute of Physics, 1995).
[46]. M. L. Heltberg, and M. H. Jensen, "Locked body clocks," Nature Physics **2019**, 15, 989-990.
[47]. M. H. Jensen, P. Bak, and T. Bohr, "Complete devil's staircase, fractal dimension, and universality of mode-locking structure in the circle map," Physical review letters **1983**, 50, 1637.
[48]. P. Bak, "Devil's staircase," Physics Today **1986**, 39, 38-45.
[49]. S. E. Brown, G. Mozurkewich, and G. Grüner, "Subharmonic Shapiro steps and devil's-staircase behavior in driven charge-density-wave systems," Physical review letters **1984**, 52, 2277.
[50]. D. Baums, W. Elsässer, and E. O. Göbel, "Farey tree and devil's staircase of a modulated external-cavity semiconductor laser," Physical review letters **1989**, 63, 155.
[51]. S. Planas, N. P. Mansur, C. B. Cruz, and H. Fragnito, "Spectral narrowing in the propagation of chirped pulses in single-mode fibers," Optics letters **1993**, 18, 699-701.
[52]. G. P. Agrawal, *Nonlinear fiber optics* (Academic press, 2007).
[53]. C. Hönninger, R. Paschotta, F. Morier-Genoud, M. Moser, and U. Keller, "Q-switching stability limits of continuous-wave passive mode locking," J. Opt. Soc. Am. B **1999**, 16, 46-56.
[54]. X. Liu, D. Popa, and N. Akhmediev, "Revealing the transition dynamics from Q switching to mode locking in a soliton laser," Physical review letters **2019**, 123, 093901.
[55]. J. Peng, M. Sorokina, S. Sugavanam, N. Tarasov, D. V. Churkin, S. K. Turitsyn, and H. Zeng, "Real-time observation of dissipative soliton formation in nonlinear polarization rotation mode-locked fibre lasers," Communications Physics **2018**, 1, 20.
[56]. K. Nozaki, and N. Bekki, "Chaos in a Perturbed Nonlinear Schr\"odinger Equation," Physical Review Letters **1983**, 50, 1226-1229.
[57]. K. J. Blow, and N. J. Doran, "Global and Local Chaos in the Pumped Nonlinear Schr\"odinger Equation," Physical Review Letters **1984**, 52, 526-529.

[58]. J. M. Soto-Crespo, and N. Akhmediev, "Soliton as strange attractor: nonlinear synchronization and chaos," Physical review letters **2005**, 95, 024101.

[59]. G. Moille, S. K. Sridhar, P. Shandilya, A. Dutt, C. Menyuk, and K. Srinivasan, "Toward Chaotic Group Velocity Hopping of an On-Chip Dissipative Kerr Soliton," Physical Review Letters **2025**, 135, 133802.

[60]. O. Rossler, "An equation for hyperchaos," Physics Letters A **1979**, 71, 155-157.

[61]. I. Fischer, O. Hess, W. Elsäßer, and E. Göbel, "High-dimensional chaotic dynamics of an external cavity semiconductor laser," Physical review letters **1994**, 73, 2188.

[62]. L. Halef, and I. Shomroni, "Route to Hyperchaos in Quadratic Optomechanics," Physical Review Letters **2025**, 135, 257201.

# Supplementary Material:
# A unified model for breathing solitons in fibre lasers: Mechanisms across below- and above-threshold regimes

Ying Zhang,[1,†] Bo Yuan,[1,†] Junsong Peng,[1,2,3,*] Xiuqi Wu,[1] Yulin Sheng,[1] Yuxuan Ren,[1] Christophe Finot,[4] Sonia Boscolo,[5,6,7] Heping Zeng[1,3,**]

[1]State Key Laboratory of Precision Spectroscopy, and Hainan Institute, East China Normal University, Shanghai 200062, China

[2]Collaborative Innovation Center of Extreme Optics, Shanxi University, Taiyuan, Shanxi 030006, China

[3]Chongqing Key Laboratory of Precision Optics, Chongqing Institute of East China Normal University, Chongqing 401120, China

[4]Université Bourgogne Europe, CNRS, Laboratoire Interdisciplinaire Carnot de Bourgogne (ICB), UMR 6303, F-21000 Dijon, France

[5]VPIphotonics, 10587 Berlin, Germany

[6]Aston Institute of Photonic Technologies, Aston University, Birmingham B4 7ET, United Kingdom

[7]Istituto Nazionale di Fisica Nucleare, Laboratori Nazionali del Sud (INFN-LNS), 95125 Catania, Italy

† These authors contributed equally to this work

*jspeng@lps.ecnu.edu.cn ** hpzeng@phy.ecnu.edu.cn

## Parameters used in the numerical model

Table 1. Normal-dispersion laser parameters

| EDF (OFS-80) | SMF (SMF-28 and HI-1060) |
|---|---|
| $L = 1.33$ m | $L = 4.16$ m |
| $\beta_2 = 61.2$ ps²/km | $\beta_2 = -21.4$ ps²/km |
| $\gamma = 3.8$ (km × W)$^{-1}$ | $\gamma = 2.0$ (km × W)$^{-1}$ |
| $\alpha = 0$ dB/km | $\alpha = 0$ dB/km |
| BW (FWHM) = 50 nm | DCF (DCF-38) |
| $\lambda_p = 980$ nm | $L = 0.53$ m |
| $\Gamma_p = 0.87$ | $\beta_2 = 65$ ps²/km |
| $\sigma_p^e = 0.00 \times 10^{-31}$ km² | $\gamma = 2.0$ (km × W)$^{-1}$ |
| $\sigma_p^a = 15.0 \times 10^{-31}$ km² | $\alpha = 0$ dB/km |
| $\lambda_s = 1550$ nm | Saturable absorber |
| $\Gamma_s = 0.87$ | $T_0 = 0.15$ |
| $\sigma_s^e = 15.0 \times 10^{-31}$ km² | $\Delta T = 0.35$ |
| $\sigma_s^a = 0.42 \times 10^{-31}$ km² | $P_{\text{sat}} = 1.25$ W |
| $A_{\text{co}} = 0.4 \times 10^{-17}$ km² | Output coupler |
| $\tau_2 = 12$ ms | $T_{\text{oc}} = 0.12$ |
| $N = 300 \times 10^{33}$ km$^{-3}$ | |

Table 2. Near-zero-dispersion laser parameters

| EDF (OFS-80) | SMF (SMF-28 and HI-1060) |
|---|---|
| $L$ = 1.25 m | $L$ = 4.30 m |
| $\beta_2$ = 61.2 ps²/km | $\beta_2$ = -16.22 ps²/km |
| $\gamma$ = 3.8 (km × W)⁻¹ | $\gamma$ = 2.0 (km × W)⁻¹ |
| $\alpha$ = 0 dB/km | $\alpha$ = 0 dB/km |
| BW (FWHM) = 50 nm | Saturable absorber |
| $\lambda_p$ = 980 nm | |
| $\Gamma_p$ = 0.87 | $T_0$ = 0.40 |
| $\sigma_p^e$ = 0.00 × 10⁻³¹ km² | $\Delta T$ = 0.13 |
| $\sigma_p^a$ = 15.0 × 10⁻³¹ km² | $P_{sat}$ = 15 W |
| $\lambda_s$ = 1550 nm | Output coupler |
| $\Gamma_s$ = 0.87 | $T_{oc}$ = 0.10 |
| $\sigma_s^e$ = 15.0 × 10⁻³¹ km² | |
| $\sigma_s^a$ = 0.42 × 10⁻³¹ km² | |
| $A_{co}$ = 0.4 × 10⁻¹⁷ km² | |
| $\tau_2$ = 12 ms | |
| $N$ = 300 × 10³³ km⁻³ | |

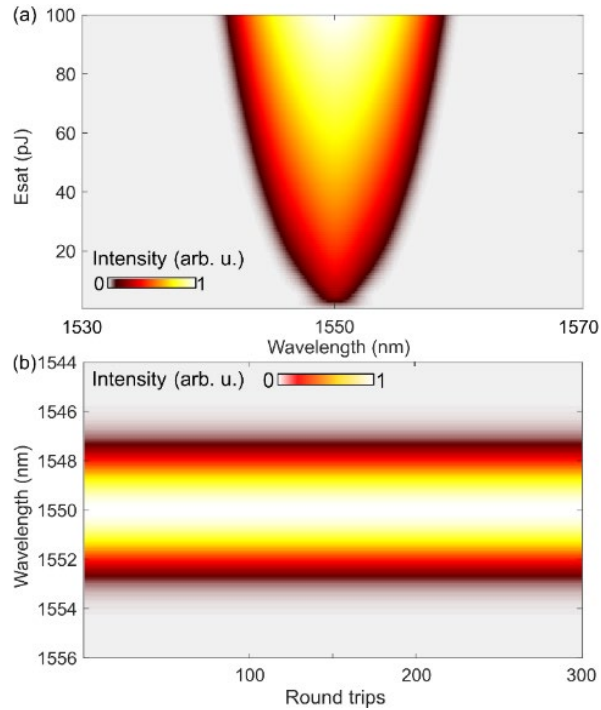

Fig. S1 Simulation of the normal-dispersion fibre laser using the standard discrete model with parameters similar to those of the modified model. The standard model fails to reproduce the breathing dynamics. (a) Pulse spectra as a function of gain saturation energy scaled from 1 to 100 pJ, showing spectral broadening but no breathing oscillations. (b) Representative roundtrip-resolved evolution of the stable spectrum at a gain saturation energy of 1 pJ.

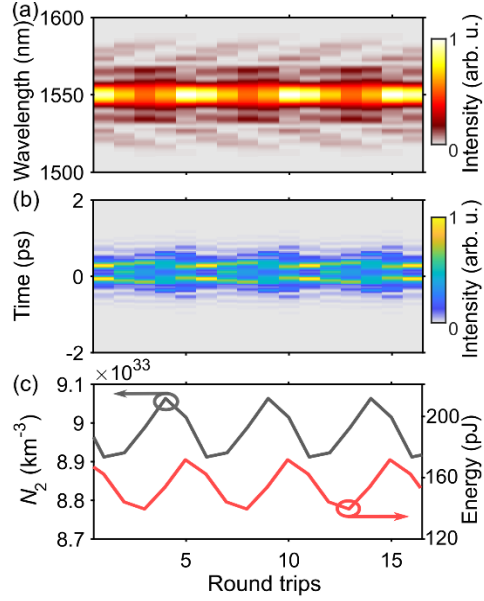

Fig. S2 Evolution of the (a) spectral and (b) temporal intensity profiles over consecutive cavity round trips for the above-threshold breathing soliton with a winding number of 1/5 depicted in Fig. 2 of the main text. As discussed therein, enhanced Kerr-dispersion interplay prevents the pulse from repeating within a single round trip. Instead, multiple round trips are required for the field to reproduce itself, with the exact number depending on the nonlinear strength. (c) Corresponding evolution of the excited-state population density ($N_2$, black) and pulse energy (red). As expected, the energy evolution lags behind $N_2$.

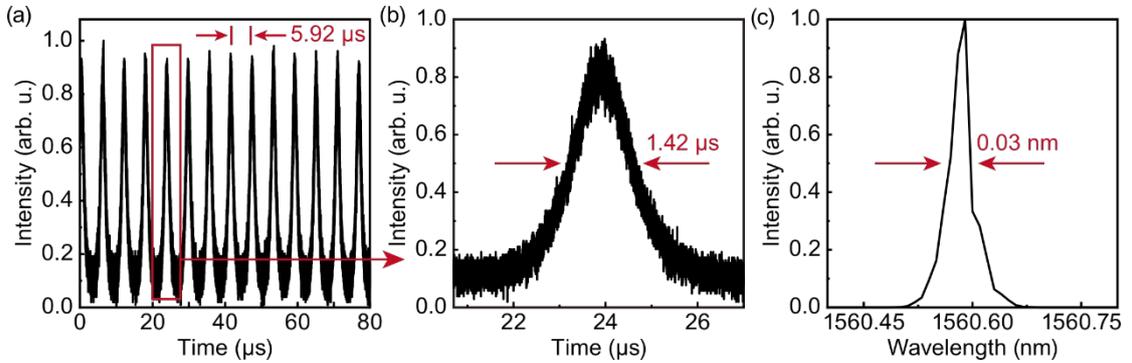

Fig. S3 Experimental observation of the Q-switching regime. (a) Q-switched pulse train with a repetition rate of 168 kHz. (b) Representative single Q-switched pulse with a duration of ~1.42 $\mu$s. (c) Corresponding optical spectrum with a 3-dB bandwidth of 0.03 nm, close to the resolution limit of the optical spectrum analyser (AQ6370D, 0.02 nm).

**Onset of Q-switching controlled by the saturable absorber**
The condition for stable mode locking (with Q-switching occurring when this condition is not satisfied) can be expressed as[1]:

$$E_p^2 > E_{\text{sat},L}\, E_{\text{sat},A}\, \Delta R, \qquad (1)$$

where $E_p$, $E_{\text{sat},L}$, $E_{\text{sat},A}$, and $\Delta R$ denote the pulse energy (which can be varied via the pump power), the saturation energy of the gain, the saturation energy of the saturable absorber, and the modulation depth of the saturable absorber, respectively. The results presented in the main text [Fig. 4(a–c)] demonstrate that the appearance of Q-switching depends on the pump power, in agreement with Eq. (1). Equation (1) further indicates that Q-switching can be suppressed by employing a smaller saturation energy ($E_{\text{sat},A}$) and/or a reduced modulation depth ($\Delta R$). As an illustrative example, Fig. S4 shows that, by decreasing only the modulation depth of the saturable absorber, Q-switching is suppressed, and stable soliton operation is achieved.

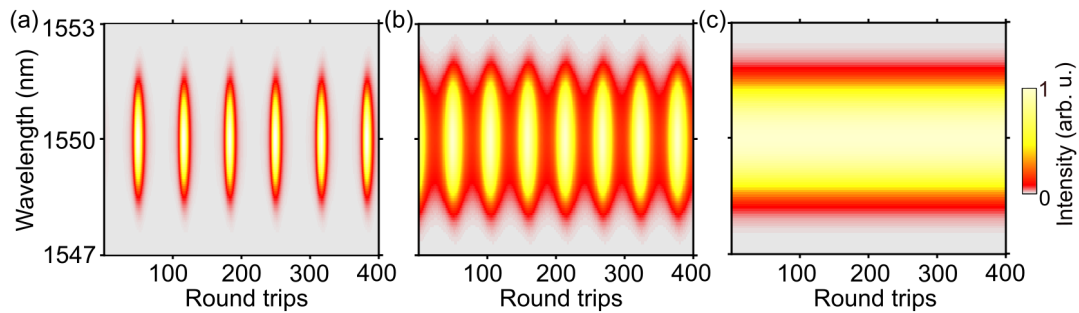

Fig. S4 (a–c) Roundtrip-resolved spectral evolution of Q-switched pulses, below-threshold breathing solitons, and stable solitons, with modulation depths of 0.15, 0.09, and 0.07, respectively. As seen, a small modulation depth (0.07) effectively suppresses Q-switching. Note that only the modulation depth is varied in these simulations.

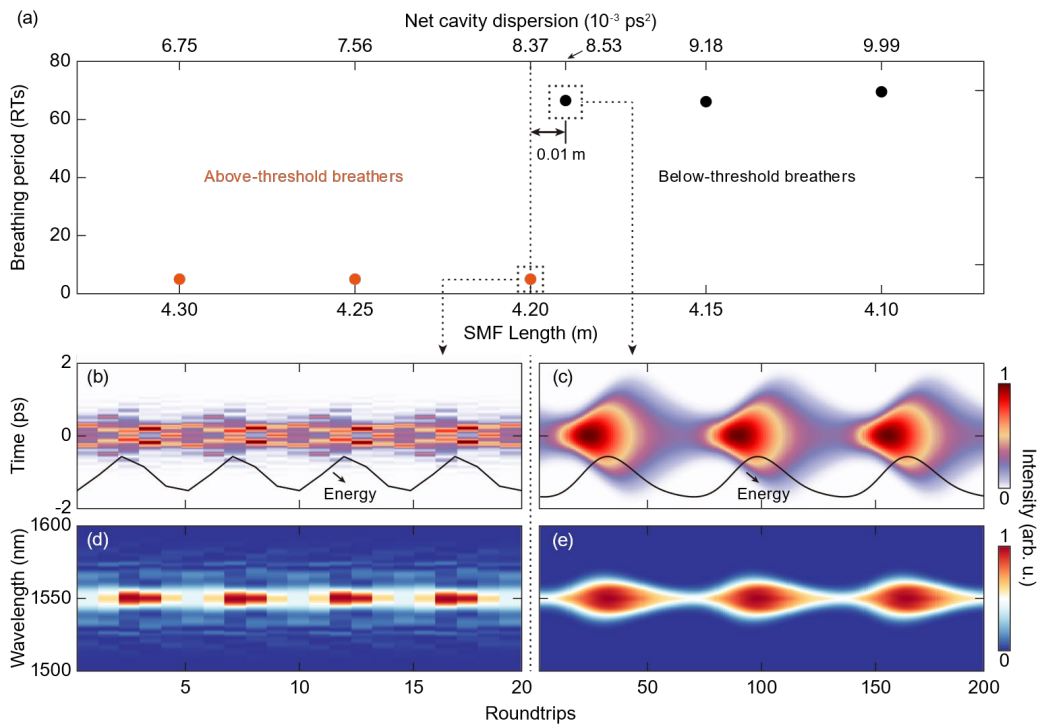

Fig. S5 Sharp transition between above- and below-threshold breathers as the net dispersion is varied by changing the length of the SMF. (a) Simulated breathing period of above- and below-threshold breathers as the SMF length is decreased from 4.30 m to 4.10 m. (b–e) Detailed temporal and spectral evolution at SMF lengths of 4.20 m and 4.19 m. Above- and below-threshold breathers can be distinguished by their breathing periods, with the former exhibiting short periods (5 round trips) and the latter much longer ones. Meanwhile, the corresponding temporal and spectral intensity profiles [(b,d) and (c,e)] are also markedly different, as discussed in the main text. The sharp transition is evident in panel (a): above-threshold breathers persist down to an SMF length of 4.20 m, while below-threshold breathers appear immediately when the SMF length is reduced by an additional 0.01 m (to 4.19 m). Note that 0.01 m corresponds to the step size used in the numerical solution of the GNLSE.

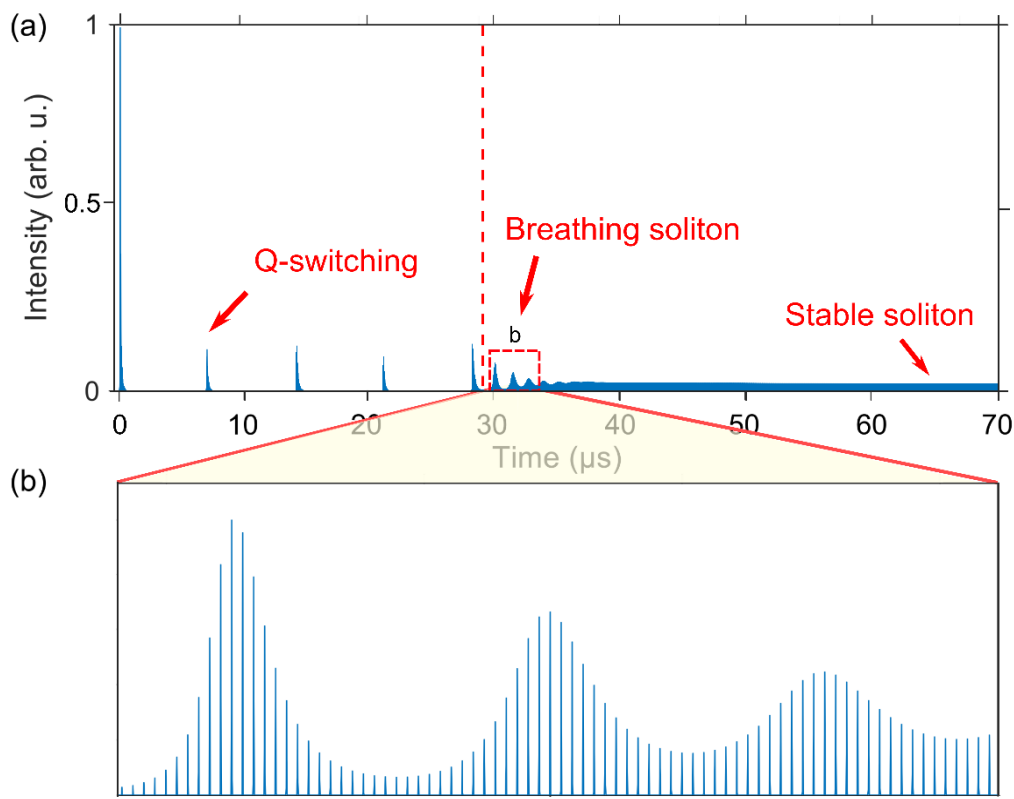

Fig. S6 Numerical simulation of the soliton buildup process observed experimentally in Ref. [2]. The model reproduces the crucial Q-switching stage preceding stable mode locking. (a) Pulse energy evolution before the formation of stable solitons, showing transient Q-switched pulses. (b) Magnified view highlighting the emergence of breathing solitons after Q-switching, consistent with the experimental observations in Ref. 1. It is important to distinguish between the usual Q-switching regime, characterised by long, low-repetition pulses, and Q-switched mode-locking, in which ultrashort mode-locked pulses are emitted in bursts modulated by slower Q-switching of the gain. The parameters used in the simulations are the same as the experiments.

**References**
[1]   C. Hönninger, R. Paschotta, F. Morier-Genoud, M. Moser, and U. Keller, "Q-switching stability limits of continuous-wave passive mode locking," J. Opt. Soc. Am. B **1999**, 16, 46-56.
[2]   P. Ryczkowski, M. Närhi, C. Billet, J. M. Merolla, G. Genty, and J. M. Dudley, "Real-time full-field characterization of transient dissipative soliton dynamics in a mode-locked laser," Nature Photonics **2018** 12, 221-227.